\documentclass[prl,reprint,sort&compress,aps,twocolumn,prb,superscriptaddress]{revtex4-2}
\usepackage{amssymb}
\usepackage{appendix}
\usepackage{verbatim}
\usepackage{float}
\usepackage{color}
\usepackage{textcomp}
\usepackage{gensymb}
\usepackage{hyperref}
\usepackage{dsfont} 
\hypersetup{
     colorlinks   = true,
     citecolor    = blue
}
\usepackage{siunitx}
\usepackage{graphicx}
\usepackage{dcolumn}
\usepackage{bm}
\usepackage{braket}
\usepackage{empheq}
\usepackage{nicefrac} 
\usepackage[dvipsnames]{xcolor}

\usepackage[normalem]{ulem}
\usepackage[vietnamese, english]{babel}

\begin{document}

\title{Interaction-Enhanced Topological  Hall Effects in Strained Twisted Bilayer Graphene}

\author{Pierre A. Pantale\'on}
\email{ppantaleon@uabc.edu.mx}
\affiliation{IMDEA Nanoscience, Faraday 9, 28049 Madrid, Spain}
\author{\foreignlanguage{vietnamese}{Võ Tiến Phong}}
\email{vophong@sas.upenn.edu}
\affiliation{Department of Physics and Astronomy, University of Pennsylvania, Philadelphia PA 19104, USA}
\author{Gerardo G. Naumis}
\affiliation{Departamento de Sistemas Complejos, Instituto de F\'isica,
Universidad Nacional Aut\'onoma de M\'exico, Apdo. Postal 20-364, 01000, Ciudad de M\'exico, CDMX, M\'exico}
\author{Francisco Guinea}
\affiliation{Imdea Nanoscience, Faraday 9, 28047 Madrid, Spain}
\affiliation{Donostia International Physics Center, Paseo Manuel de Lardiz\'abal 4, 20018 San Sebastián, Spain and Ikerbasque, Basque Foundation for Science, 48009 Bilbao, Spain}

\begin{abstract}

We analyze the effects of the long-range Coulomb interaction on the distribution of Berry curvature among the bands near charge neutrality of twisted bilayer graphene (TBG) closely aligned with hexagonal boron nitride (hBN). Due to the suppressed dispersion of the narrow bands, the band structure is strongly renormalized by electron-electron interactions, and thus, the associated topological properties of the bands are sensitive to filling. Using a Hartree formalism, we calculate the linear and nonlinear Hall conductivities, and find that for certain fillings, the remote bands contribute substantially to the Hall currents while the contribution from the central bands is suppressed. In particular, we find that  these  currents are generically substantial near regions of energies where the bands are highly entangled with each other, often featuring doping-induced band inversions. Our results demonstrate that topological transport in TBG/hBN is substantially modified by electron-electron interactions, which offer a simple explanation to recent experimental results.

\end{abstract}

\maketitle

{\it Introduction \textendash} Transverse Hall currents are usually associated with broken time-reversal symmetry. However, Hall-like currents are possible in time-reversal symmetric settings with broken inversion symmetry at second order in the applied  in-plane electric field~\citep{TonyLowPaco2015,SodemannLiang2015,NandySodemman2019,Du2021Rev}. This so-called nonlinear Hall effect has been observed in transition-metal dichalcogenides (TMDs)~\citep{Ma2019,KangMak2019,huang2020giant,hu2020nonlinear} and other two dimensional materials~\citep{Araki2018, XiaoHua2020, FacioJeroen2018,Wang2019c,Du2019,Liao2021AI,Joseph2021Janus}. Furthermore, it has been recently shown that uniaxial strains can enhance the Berry dipole~\citep{YouLow2018,ZhouZhangLaw2019,SonKimLee2019}, defined as the gradient of the Berry curvature, in TMD that is responsible for the nonlinear Hall effect. For these materials, orbital valley magnetization~\citep{SonKimLee2019, ShiJustinMagnet2019}, giant magneto-optical effects~\citep{Liu2020}, and nonlinear Nernst effects~\citep{YuTony2019,ZengTewari2019}  can also be induced with an applied field due to the sizeable Berry dipole.  

Recent theoretical works have shown that nonlinear Hall effects can also arise in gapped and strained twisted bilayer graphene (TBG)~\citep{Pantaleon2021Dipole,Zhang2020GiantNTBG,Arora2021Inj}, which depend sensitively on filling, and can be larger than in TMDs. The magnitude of the Berry dipole is predicted to be larger near half filling of the narrow conduction and valence  bands. However, a recent experiment observes nonlinear Hall currents in TBG on hexagonal boron nitride (hBN) with negligible contribution from the narrow bands~\citep{Duan2022Nonlin}. On the contrary, large nonlinear Hall currents are observed when the chemical potential is tuned to a spectral region entangled with the remote bands. On the basis of their experimental results, the authors of Ref.~\citep{Duan2022Nonlin} suggest that the nonlinear Hall effect is due primarily to skew scattering from dynamic disorders~\citep{Du2019} instead of a pure Berry dipole induced by a heteroaxial strain and a mass gap \footnote{In fact, the authors of Ref.~\citep{Duan2022Nonlin} do not consider strain. So their system retains some twofold rotations that force the Berry dipole to vanish. They justify this exclusion by arguing that the contribution from strain is much smaller than the extrinsic contributions.}. In addition to TBG, nonlinear Hall currents have also been measured in the related system of strained twisted double bilayer graphene (TDBG)~\citep{Sinha2022diposenses}. This experiment in particular finds that the \textit{sign} of the Berry dipole is sensitive to band inversions, and therefore is a signature of topological band transitions \footnote{In TDBG, these band inversions can be induced by applying an interlayer bias voltage. When the bias is large enough, the active middle bands can hybridize with the remote bands and create a sign change in the local Berry curvature}.

Inspired by these experiments, we theoretically study strained TBG in the presence of an hBN substrate. Our results indicate that the Berry dipole contributes significantly to the large nonlinear Hall effect found in the experiment of Ref.~\cite{Duan2022Nonlin} and is strongly enhanced near avoided band crossings similar to the behavior of TDBG in Ref.~\citep{Sinha2022diposenses}. Without invoking any extrinsic influence from disorder, we find that the combination of a small uniaxial heterostrain, an hBN substrate, and electron-electron interactions treated within a self-consistent Hartree formalism are enough to explain the large observed nonlinear Hall currents derived from the remote bands~\cite{Duan2022Nonlin}. In particular, we find that the Coulomb interaction simultaneously distorts the energy spectrum and redistributes the Berry curvature, both of which can lead to large nonlinear Hall currents. Our work demonstrates that because of significant interaction-meditated band distortions, the strong Coulomb repulsion in twistronic systems is not only important for inherently many-body phenomena like magnetism and superconductivity, but is also crucial for determining physics that is often otherwise studied from a single-particle perspective.

\begin{figure}
\begin{centering}
\includegraphics[scale=0.75]{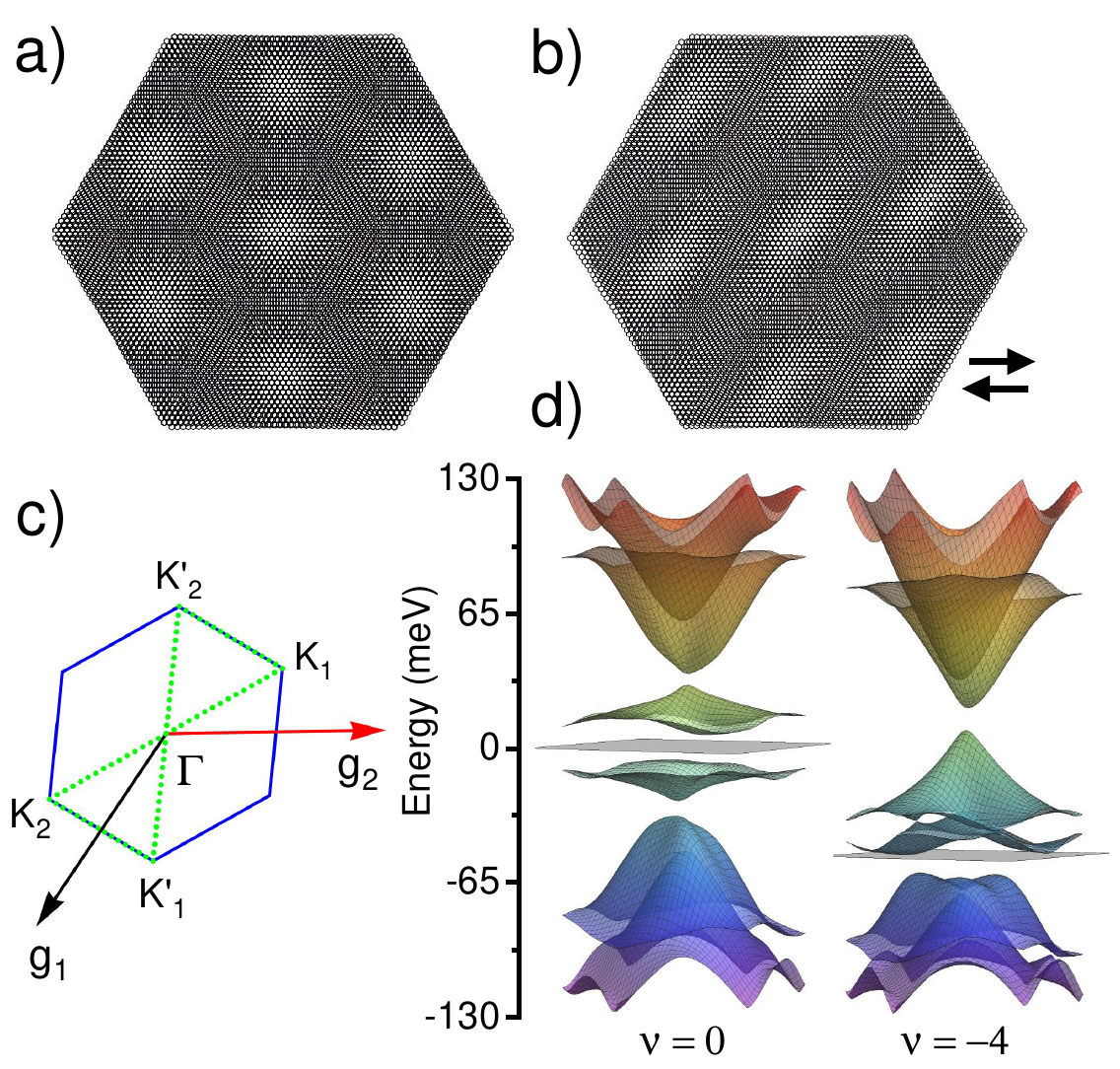}
\par\end{centering}
\caption{(a) Schematic of the moir\'e superlattice for unstrained TBG and (b) heteroaxially strained TBG, where for visual clarity, we set $\epsilon =3 \% $ for the relative strain magnitude and $\phi=0$ for the strain direction. We are assuming that the moir\'e pattern of the hBN substrate is the same as  TBG (see Ref.~\cite{SI} for details). In (b), black arrows indicate the heterostrain directions. (c) Distorted moir\'e Brillouin zone where the reciprocal lattice vectors $\boldsymbol{g}_{1,2}$ are indicated. Green dotted path is used to calculate the band structure. (d) Three-dimensional plots of the band structure showing six central bands for strained TBG with $\epsilon =0.2 \% $ and $\phi=0$ suspended on hBN and for  filling factors $\nu=0$ (left) and  $\nu=-4$ (right). The Fermi levels are indicated by the square shaded meshes.}
\label{fig: Figure1}
\end{figure}

\begin{figure*}
\begin{centering}
\includegraphics[scale=0.27]{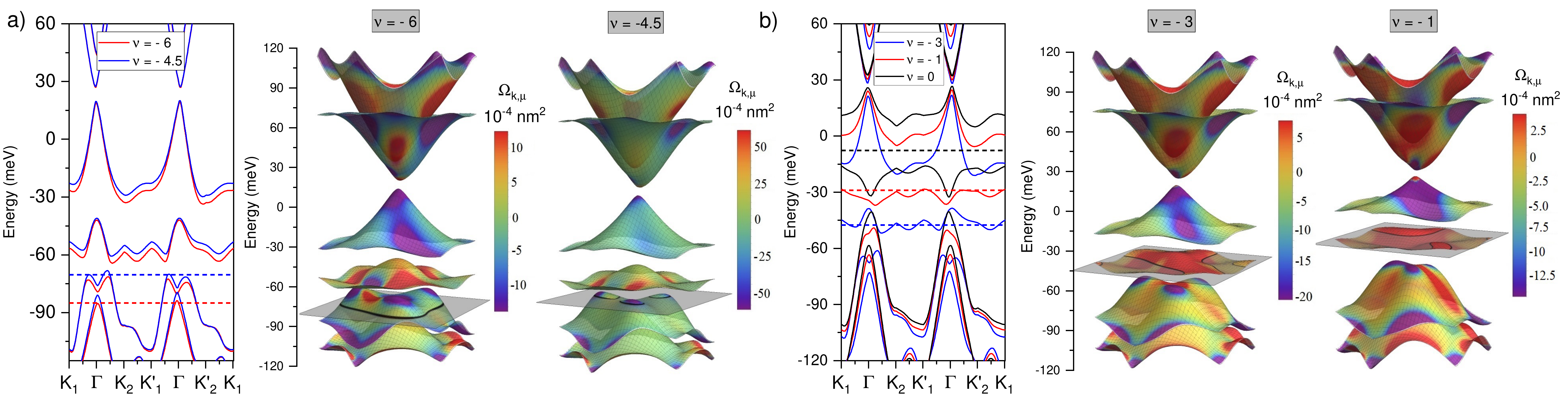}
\par\end{centering}
\caption{Band structure of strained TBG/hBN for twist angle $\theta=1.05^\circ$ and for different filling factors. (a) For a Fermi level within the remote bands, there are massive Dirac cones (with small masses) that significantly contribute to the Berry curvature. In (b), we show the bands for fillings  within the lower narrow band. The dashed horizontal lines in each plot are the positions of the Fermi level. Adjacent to each plot, we display the full band structure for the corresponding filling. The color texture in each band is the magnitude of the Berry curvature as indicated in the color bar. Gray square surfaces indicate the positions of the chemical potential and black contours their intersection with the energy bands. In all plots, we use a relative strain magnitude $\epsilon=0.2 \%$ with direction $\phi=0$.
\label{fig: Figure2}}
\end{figure*}

 {\it Symmetries in strained TBG/hBN \textendash} In pristine TBG graphene, both time-reversal and inversion symmetries are preserved, as shown in Fig.~\ref{fig: Figure1}(a). For small twist angles, there are two narrow bands near charge neutrality that are connected by two Dirac points at the corners of the moir\'{e} Brillouin zone (mBZ). These Dirac points are protected by a robust $\mathcal{C}_{2z}\mathcal{T}$ symmetry, where $\mathcal{C}_{2z}$ is twofold rotation about the $z$-axis and $\mathcal{T}$ is time reversal~\footnote{ We note that although both $\mathcal{C}_{2z}$ and $\mathcal{T}$ map one valley to the other,  the combination $\mathcal{C}_{2z}\mathcal{T}$ preserves the valley index.}. To gap out these Dirac points and obtain finite Berry curvature in the narrow bands, we need to break $\mathcal{C}_{2z}\mathcal{T}$ by placing TBG on hBN~\cite{CeaPantaGuinea2020,Zhang2019b,ShiComm2021,Sharpe2021Orbital,Huang2021Imag,Lin2020Sym,Lin2021Mis}. The inversion-broken character of hBN due to sublattice inequivalence is transmitted to the graphene bands by means of scalar, mass, and strain potentials~\cite{Wallbank2013, San-Jose2014, Mucha-Kruczynski2013,Jung2017}. While the magnitude of these inversion-broken gaps are extremely sensitive to the degree of alignment between TBG and hBN, they are non-zero even for a large misalignment~\citep{LongJose2021}. 
 
Another important feature of TBG that was experimentally determined by STM measurements is the presence of a strain field that alternates sign between the two layers (uniaxial heterostrain) and a magnitude between $0.1$ and $0.5\%$~\cite{Jetal19,Choi2019,Xie2019,Kerelski2019,Quiao2018ht,Huder}. Such a strain field further breaks  $\mathcal{C}_{3z},$ $\mathcal{C}_{2x},$ and $\mathcal{C}_{2y}$ rotation symmetries. However, $\mathcal{C}_{2z}$ remains intact under strain, as demonstrated in  Fig.~\ref{fig: Figure1}(b). In momentum space, strain distorts the mBZ as shown in  Fig.~\ref{fig: Figure1}(c). The Dirac cones of the distorted lattice are no longer at the same energy or at the corners of the mBZ~\citep{OlivaLeyva2015,BiFu2019,Parker2021Str}. When both valleys are considered, TBG both with and without strain respect $\mathcal{T}$ symmetry. Therefore, only the nonlinear Hall effect can  survive since the linear Hall term is odd under time reversal. In what follows, when we calculate the linear term, it will always be for a single valley, which can be nonzero since each valley individually breaks $\mathcal{T}.$ In a transport experiment, both valleys are measured simultaneously; so the leading contribution to the Hall currents comes at second order in the applied electric field.  

{\it Coulomb interactions in strained TBG/hBN \textendash} In transport experiments on TBG, the Fermi level can be tuned. Due to long-range  electron-electron interactions, it has been theoretically predicted~\citep{Cea2019,Guinea2018a,Rademaker2019} and experimentally found~\citep{Xie2019,Jiang2019a,Tomarken2019,Xiaobo_nat19} that the Fermi level is \textquotedblleft pinned\textquotedblright to the van Hove singularities. Away from charge neutrality, it has been shown that the Coulomb interaction is comparable or even larger than the bandwidth of the central bands, giving rise to a filling-dependent Hartree potential that substantially modifies the shape of the electronic bands~\citep{Guinea2018a,Rademaker2019}. In strained TBG/hBN with a Hartree interaction, the band structure strongly depends on the position of the Fermi level, as shown in Fig.~\ref{fig: Figure1}(d) and  Fig.~\ref{fig: Figure2}. This implies that any physical quantity involving a variation of the chemical potential near or within the narrow bands in pristine TBG or TBG/hBN should account for the variation of the Hartree potential.

\begin{figure*}
\begin{centering}
\includegraphics[scale=0.44]{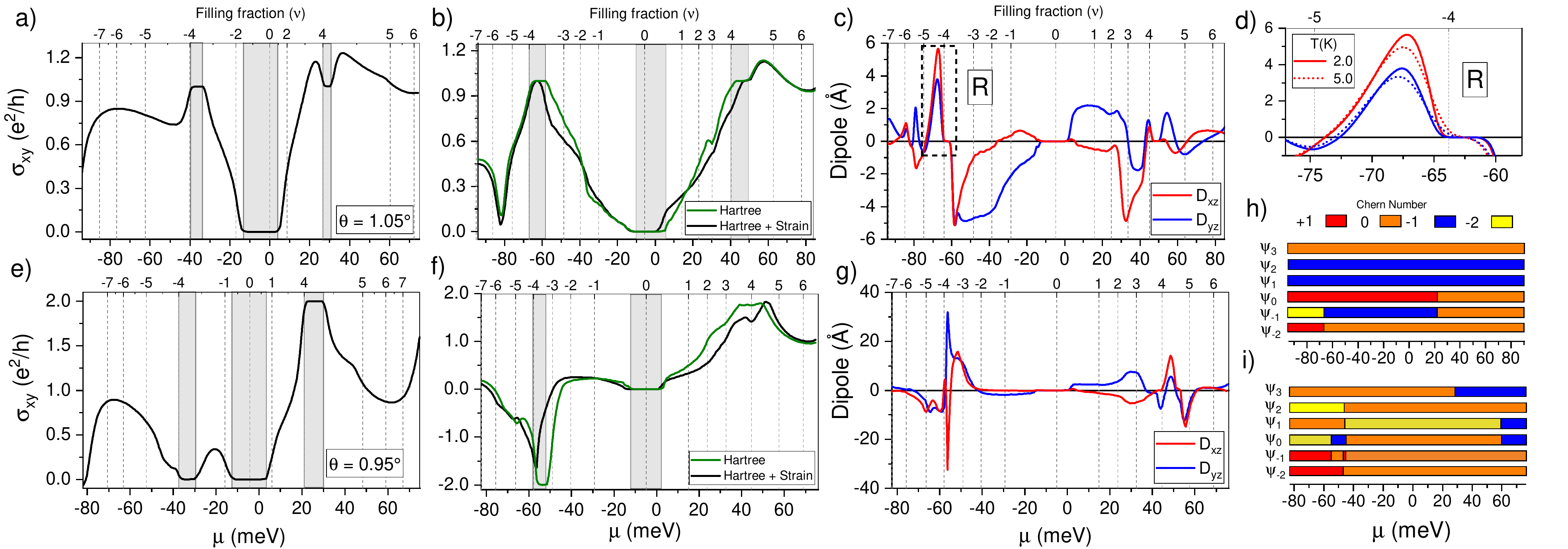}
\par\end{centering}
\caption{Interaction-enhanced topological Hall effects in TBG/hBN. In the top row, for a twist angle of $\theta = 1.05^\circ,$ we plot the valley Hall conductivity as a function of  chemical potential for (a) the rigid system and (b) with interactions and strain. Gray shaded areas are gapped regions without strain. In (c), we display the Berry dipole as a function of $\mu$. In (d), we show an enlarged region of (c), labelled as R, where the dependence on temperature is indicated. Bottom row is for a twist angle $\theta = 0.95^\circ$, where we plot the valley Hall conductivity for (e) the rigid system and (f) with interactions and strain. (g) Berry dipole  featuring a  prominent peak near $\nu = -4.$ Panels (h) and (i) display the Chern numbers of the six central bands, with $\psi_{i}$, where $i=-2,...,3$, labeling the bands from lower to higher energy, as a function of $\mu$ for (h) $\theta = 1.05^{\circ}$ and (i) $\theta = 0.95^{\circ}$. Several doping-induced band inversions are observed. In all  plots, unless otherwise indicated, we use $T=5$ K, $\epsilon_{d} =10$ for the dielectric constant, and $\epsilon = 2\%$ and $\phi=0$ for strain. All calculations are done for one spin species in a single valley. 
\label{fig: Figure3}}
\end{figure*}

\begin{figure}
\begin{centering}
\includegraphics[scale=0.49]{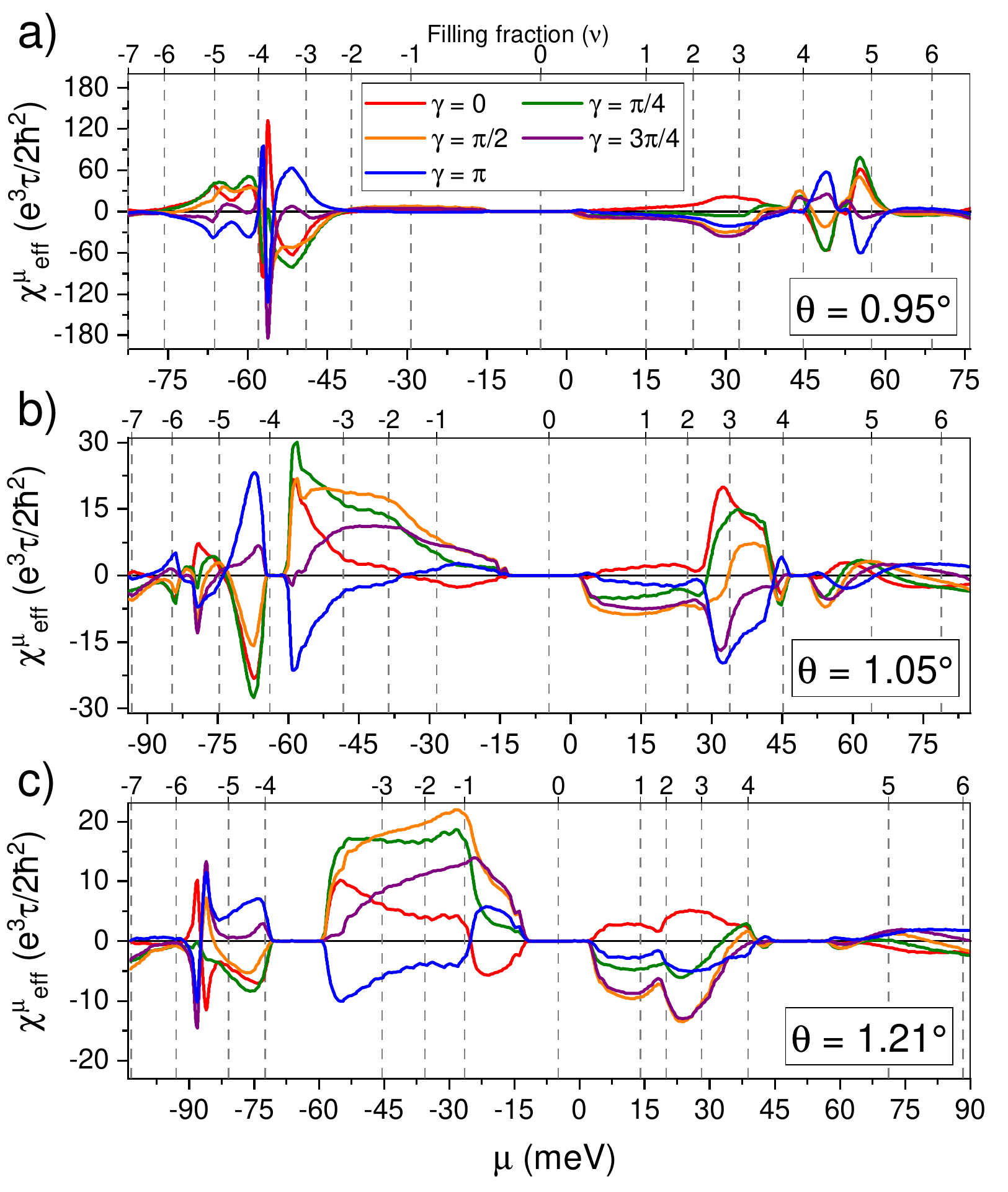}
\par\end{centering}
\caption{Effective nonlinear susceptibility in strained TBG/hBN for different current directions with (a) $\theta = 0.95^{\circ}$, (b) $\theta = 1.05^{\circ}$, and (c) $\theta = 1.21^{\circ}$. For small twist angles, the Berry dipole is strongly enhanced near $\nu = \pm 4.$   
\label{fig: Figure4}}
\end{figure}

{\it Linear and nonlinear Hall currents \textendash} With the preceding considerations, we now calculate the topological currents arising from the Berry curvature with electron-electron interaction encoded in a self-consistent Hartree potential.   In this setting, within the semiclassical transport theory in the presence of an AC electric field $\mathcal{E}(t)=\text{Re}\{\boldsymbol{\xi}e^{i\omega t}\}$ with frequency $\omega$, the band velocity is given by     $\hbar\boldsymbol{v}_{n}(\boldsymbol{k})=\nabla_{k}E_{n}(\boldsymbol{k})+e\mathcal{E}\times\boldsymbol{\Omega}_{n}(\boldsymbol{k})$, where the first term is the usual group velocity and the second is an anomalous term driven by a nonzero Berry curvature~\citep{XiaoBerry2010}. For us, $\boldsymbol{\xi}$ is a real vector. The band velocity $\boldsymbol{v}_{n}(\boldsymbol{k})$ has a component transverse to the electric field that gives rise to Hall currents. The Berry curvature for the electronic Bloch states of the $n^\text{th}$ band is $\Omega_{n,\mu}^{z}(\boldsymbol{k})=2 \mathrm{Im}\left\langle \partial_{k_{x}}\Psi_{n\boldsymbol{k},\mu}\right|\left.\partial_{k_{y}}\Psi_{n\boldsymbol{k},\mu}\right\rangle \hat{z}$, where we have introduced an index $\mu$ to emphasize that the Berry curvature depends on $\mu$ via the Hartree potential. Following Refs.~\citep{Du2019,du2020quantum,TonyLowPaco2015,SodemannLiang2015}, the transverse currents up to second order in the electric field are $\mathcal{J}_{a,\mu}=\text{Re}\{\mathcal{J}_{a,\mu}^0+\mathcal{J}_{a,\mu}^\omega e^{i\omega t}+\mathcal{J}_{a,\mu}^{2\omega} e^{i2\omega t}\}$ with a rectified component $\mathcal{J}_{a,\mu}^0=\chi_{abc}^{\mu}\xi_{b}\xi_{c}^*$, a linear term $\mathcal{J}_{a,\mu}^\omega=\sigma_{ab}^{\mu}\xi_{b}$, and a second-harmonic component $\mathcal{J}_{a,\mu}^{2\omega}=\chi_{abc}^{\mu}\xi_{b} \xi_{c}$. The linear term contains the usual valley Hall conductivity. The susceptibility in the first and third term is given by $\mathcal{\chi}_{abc}^\mu  =  -\frac{e^{3}\tau}{2\hbar^{2}}\varepsilon^{acd}D_{bd,\mu}$ with Berry dipole (see sec. III in Ref.~\cite{SI}),
\begin{equation}
D_{bd,\mu}=\sum_{n}\int \frac{d\boldsymbol{k}}{(2\pi)^{2}} \Omega_{n,\mu}^{d}(\boldsymbol{k})\partial_{b}f_{0}[\epsilon_{n,\mu}(\boldsymbol{k})]
\label{eq: Berry Dipole Main}.
\end{equation}
Notice that contrary to a system with rigid bands~\citep{Pantaleon2021Dipole,Zhang2020GiantNTBG}, the wavefunctions required to obtain the Berry curvature must be calculated self-consistently for each value of the Fermi energy, as shown in Fig.~\ref{fig: Figure2}.  The Berry dipole depends on the chemical potential in two different ways. The first is an explicit dependence in $f_0$ in the integral. The second is an \textit{implicit} dependence due to the redistribution of the Berry curvature as charges reorganize due electrostatic interactions when electrons are added to or removed from the system. In rigid platforms where these electrostatic interactions can be neglected like in TMDs, the implicit dependence can be ignored. However, interestingly in our case, the redistribution of the Berry curvature when $\mu$ varies plays a crucial role in determining the magnitude of the nonlinear Hall effect.

{\it Interaction-enhanced topological Hall effects \textendash} Fig.~\ref{fig: Figure3} summarizes the topological Hall effects in strained TBG/hBN as a function of $\mu$ for an energy window covering the six central bands. Fig.~\ref{fig: Figure3}(a) and (e) show the valley Hall conductivity for the rigid system (TBG/hBN without interactions or strain) at $\theta = 1.05^\circ$ and $\theta = 0.95^\circ$ respectively. Fig.~\ref{fig: Figure3}(b) and (f) show the corresponding Hall conductivity with interactions; there, the behavior is especially different compared to the rigid system near $\nu = \pm 4$. For instance, the Hall conductivity at $\nu = -4$ and $\theta = 0.95^\circ$ vanishes when interactions are neglected, but is about $-2e^2/h$ when interactions are included. In addition, the range of chemical potential that resides in the central bands is significantly widened by the Hartree potential~\citep{Cea2019}. We also note that whenever the chemical potential is inside a bulk mobility gap, $\sigma_{xy}^{\mu}$ is quantized and is given by the Chern numbers of the bands below chemical potential. The Chern numbers of the six bands closest to charge neutrality are tabulated in Fig.~\ref{fig: Figure3}(h)~and~(i).

Further straining the system in addition to the Hartree potential does not significantly change $\sigma_{xy}^{\mu}$, as shown by the black lines in Fig.~\ref{fig: Figure3}(b) and (f). Therefore, strain seems to play a minimal role in the linear valley Hall effect. On the contrary, strain is crucial for the nonlinear Hall effect since it permits a non-zero Berry dipole to develop, as shown in Fig.~\ref{fig: Figure3}(c) and (g). In contrast to earlier non-interacting calculations~\citep{Pantaleon2021Dipole,Zhang2020GiantNTBG} where the Berry dipole near magic angle is found to be large especially when the Fermi level is within the narrow bands, we find in the present work using realistic parameters that the Hartree potential enhances the Berry dipole near the edges of the narrow central bands. The Berry dipole is large when either the Berry curvature, the band velocity, or both are large near the Fermi surface. Furthermore, the Berry curvature can also be made large when the energy separation between occupied and unoccupied bands is very small.  Therefore, the Berry dipole peaks where either (i) the group velocity is large, or (ii) the energy bands are very close together. The former contribution generically occurs where an initial band crossing (such as a Dirac cone) is gapped out by some symmetry-breaking perturbations, which is the case for $\nu = -4.5$ as shown in Fig.~\ref{fig: Figure2}(a). The latter contribution comes from regions in energies where the bands are highly entangled so that the separation between occupied and unoccupied bands is quite small. This is the case for $\nu = -6$ in Fig.~\ref{fig: Figure2}(a). These two contributions are generically large near band edges, which intuitively explains why we often observe large nonlinear Hall responses for $\nu$ near $-4.$

In Fig. \ref{fig: Figure3}(c) and (g), the Berry dipole pseudovector $\boldsymbol{D}_\mu = (D_{xz,\mu},D_{yz,\mu})$ is defined relative to the lattice structure. In an experiment, it is difficult to isolate the effect of either component of the Berry dipole. Instead, an applied electric field $\boldsymbol{\xi}$ at a generic angle $\gamma$ relative to the crystallographic axes mixes these components such that the effective angle-dependent susceptibility is $\chi_\text{eff}^\mu(\gamma) = -4\frac{e^3 \tau}{2\hbar^2}  \boldsymbol{D}_\mu \cdot \hat{\boldsymbol{\xi}},$ where the prefactor $4$ accounts for spin and valley degeneracies. The effective susceptibility for various field angles $\gamma$ at different twist angles $\theta$ are plotted in Fig. \ref{fig: Figure4}. We emphasize a few interesting features. First, $\chi_\text{eff}^\mu$ is concentrated in highly-entangled spectral regions, especially for small angles. This enhancement  is sometimes made possible by \textit{doping-induced} band inversions. For instance, near $\mu = -67$ meV, we see in Fig. \ref{fig: Figure3}(h) that there is a band inversion between bands $\psi_{-2}$ and $\psi_{-1},$ resulting in a large susceptibility shown in Fig.~\ref{fig: Figure4}(b). Second, a change in the \textit{sign} of $\chi_\text{eff}^\mu$  indicates a change in the local (but not necessarily global) band geometry near the Fermi surface. For example, we see that $\chi_\text{eff}^\mu$ changes sign near $\nu = -4$ at $\theta = 1.05^\circ.$ This is precisely the energy region where the linear Hall conductivity (shown in Fig. \ref{fig: Figure3}(b)) reaches a maximum, signifying a change in the sign of the Berry curvature. This reversal in local band geometry can occur both as $\mu$ sweeps across a bulk gap, such as the case  near $\nu = -4$ at $\theta = 1.05^\circ,$ and as $\mu$ varies within a single bulk band, such as the case near $\nu = -1.8$ at $\theta = 1.05^\circ$ and $\gamma = 0 \text{ or } \pi$. Similar polarity reversals are observed for $\theta = 0.95^\circ$ and $\theta = 1.21^\circ$ as well \footnote{When there is an external tunable parameter such as an interlayer bias voltage, the sign of the Berry dipole can also be used to detect the occurrence of a band inversion driven by that parameter. This is precisely the case in Ref. \cite{Sinha2022diposenses} }.

{\it Discussion and conclusion \textendash} Our results are in qualitative agreement \footnote{Quantitatively reproducing the experimental measurements from our theory is more challenging because the scattering time is not empirically known and is difficult to calculate from first principles. In addition, there are several adjustable parameters in the setup that can complicate a quantitative comparison such as the degree of alignment between TBG and the substrate and  the direction and magnitude of strain relative to the lattice. Nonetheless, we have made some rough numerical estimates in Ref. \cite{SI} that suggest a sensible order-of-magnitude agreement.} with the recent experiment in TBG~\cite{Duan2022Nonlin} where a large nonlinear Hall current is observed when the chemical potential is tuned to a spectral region entangled with the remote bands. The authors of the experiment attribute this observation to skew scattering due to magnetic impurities. However, we can explain the large observed signals at the edges of the narrow bands by carefully accounting for band structure renormalization due to the Hartree potential and the presence of an hBN substrate without considering disorder. Our theory cannot exclude the role of disorder in the experimental results, but it at least demonstrates that the Berry dipole also contributes significantly to the Hall current at filling factors near $\pm 4$ if  hBN is nearly aligned with TBG. Importantly, the source of this Berry dipole enhancement is unquestionably Coulomb repulsion. In addition, our result is also of general agreement to the conclusion of Ref.~\citep{Sinha2022diposenses} that the Berry dipole is sensitive to band inversions, although our topological transitions are induced by doping instead of a bias. Nonetheless, this result demonstrates that although the Berry dipole is a Fermi surface property, it can sometimes be used to infer global band topology.  Encouraged by this, we believe that adopting the theoretical framework developed here to study nonlinear Hall effects in other experimentally-relevant  twisted two-dimensional materials is a promising future research direction.

{\it Acknowledgements.} IMDEA Nanociencia acknowledges support from the \textquotedblleft Severo Ochoa\textquotedblright  Programme for Centres of Excellence in R\&D (Grant No. SEV-2016-0686). P.A.P and F.G. acknowledge funding from the European Commission, within the Graphene Flagship, Core 3, grant number 881603 and from grants NMAT2D (Comunidad de Madrid, Spain), SprQuMat. VTP acknowledges support from the NSF Graduate Research Fellowships Program and the P.D. Soros Fellowship for New Americans. G.N. thanks UNAM-DGAPA project IN102620 and CONACyT project 1564464.

\bibliography{References.bib}


\clearpage
\onecolumngrid

\setcounter{equation}{0}
\setcounter{figure}{0}
\setcounter{table}{0}
\setcounter{page}{1}
\makeatletter
\renewcommand{\theequation}{S\arabic{equation}}
\renewcommand{\thefigure}{S\arabic{figure}}

\begin{center}
\Large Supplementary Material for \\
Interaction-Enhanced Topological Hall Effects in Strained Twisted Bilayer Graphene
\end{center}

\section{Uniaxial Heterostrain in TBG}

\begin{figure*}
\begin{centering}
\includegraphics[scale=0.25]{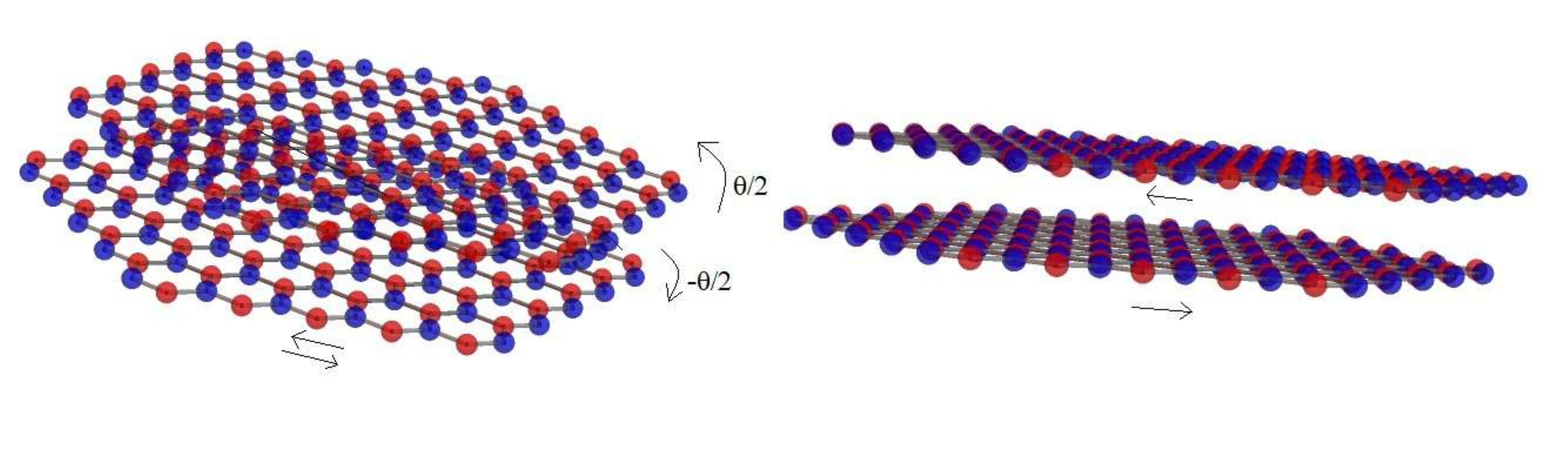}
\par\end{centering}
\caption{Lattice geometry of strained twisted bilayer graphene in two different visualizations. Twist directions are indicated by curved arrows. The black horizontal arrows show the direction of the uniaxial heterostrain. The different colors indicate atoms on each bipartite sublattice.
\label{fig: FigureGeo}}
\end{figure*}

In  monolayer graphene, the primitive lattice vectors are $\boldsymbol{a}_{1}=a(1,0)$ and $\boldsymbol{a}_{2}=a(1/2,\sqrt{3}/2)$ with lattice constant $a\approx2.46 \text{ \AA}.$  The reciprocal lattice vectors, $\boldsymbol{b}_{i}$, $i=1,2$ satisfying $\boldsymbol{a}_{i}\cdot \boldsymbol{b}_{j}=2\pi\delta_{ij},$ are given by $\boldsymbol{b}_{1}=\frac{2\pi}{a}(1,-1/\sqrt{3})$ and $\boldsymbol{b}_{2}=\frac{2\pi}{a}(0,2/\sqrt{3})$. The graphene Dirac cones are located at $\boldsymbol{K}_{\zeta}=-\zeta(2\boldsymbol{b}_{1}+\boldsymbol{b}_{2})/3 = - \zeta 4\pi/3a (1,0)$ with $\zeta=\pm1$ denoting the valley index. If $R(\theta)$ represents a rotation matrix by $\theta$, the primitive and reciprocal lattice vectors in each rotated layer $l$ are written as $\boldsymbol{a}_{i}^{(l)}=R(\mp\theta/2)\boldsymbol{a}_{i}$ and $\boldsymbol{b}_{i}^{(l)}=R(\mp\theta/2)\boldsymbol{b}_{i}$. 
Here, $\theta/2$ is the rotation angle of the corresponding layer (see Fig. \ref{fig: FigureGeo}). We now introduce a geometric uniaxial deformation, where the bilayer system is relatively stressed along one direction and unstressed along the perpendicular direction~\citep{OlivaLeyva2015,BiFu2019,Manna2021St}. Geometrically, uniaxial strain can be described by two parameters, the relative strain  magnitude $\epsilon$ and the strain direction $\phi$. The strain tensor $\mathfrak{\varepsilon}$ in terms of these two parameters is written as 
\begin{equation}
\varepsilon=\epsilon\left(\begin{array}{cc}
-\cos^{2}\phi+\nu\sin^{2}\phi & (1+\nu)\cos\phi\sin\phi\\
(1+\nu)\cos\phi\sin\phi & -\sin^{2}\phi+\nu\cos^{2}\phi
\end{array}\right),\label{eq: StrainTensor}
\end{equation}
where $\nu=0.16$ is the Poisson ratio for graphene~\cite{Faccio2009Mech}. In TBG with alternating uniaxial heterostrain, as shown in  Fig.~\ref{fig: FigureGeo}, each layer is strained in the opposite direction relative to the other layer. Thus, the transformed primitive and reciprocal lattice vectors for each rotated graphene layer are given by 
\begin{align}
\boldsymbol{\alpha}_{i}^{(l)} & =(\mathbb{I}+\varepsilon^{(l)})\boldsymbol{a}_{i}^{(l)},\label{eq: StrainRealPrimitive}\\
\boldsymbol{\beta}_{i}^{(l)} & =(\mathbb{I}-[\varepsilon^{(l)}]^{T})\boldsymbol{b}_{i}^{(l)},\nonumber 
\end{align}
with $\varepsilon^{(l)}$ the strain tensor, $l$ the layer index, and $\mathbb{I}$ a $2 \times 2$ identity matrix. The reciprocal lattice vectors of the {\it deformed moir\'e superlattice} are
\begin{equation}
\boldsymbol{g}_{i}=\boldsymbol{\beta}_{i}^{(1)}-\boldsymbol{\beta}_{i}^{(2)}.
\label{eq: Straingvectors}
\end{equation}
The relative deformation between consecutive graphene layers satisfies $\varepsilon=\varepsilon^{(2)}-\varepsilon^{(1)}$ with $\varepsilon^{(2)} = -\varepsilon^{(1)} =\frac{1}{2}\varepsilon$~\cite{BiFu2019}. We remark here that the symmetries of the lattice generated from $\boldsymbol{\alpha}_{i}^{(l)}$  are changed when compared with the unstrained lattice. Therefore, the reciprocal space lattice, the Brillouin zone, and the position of the high-symmetry points  $\bm{K}_{\zeta}^{(l)}$ are changed accordingly \cite{Oliva2013,OlivaLeyva2015,Naumis2017}. In particular, the original $\bm{K}_{\zeta}^{(l)}$ points on each layer are now shifted to positions 
\begin{equation}
    \tilde{\bm K}_{\zeta}^{(l)}=(\mathbb{I}-[\varepsilon^{(l)}]^{T})\bm{K}_{\zeta}^{(l)}.
\end{equation} 
In addition to these geometrical effects, strain also modifies the electronic structure by a change in the intralayer Hamiltonian. The origin of this effect stems from the modulation of the bond amplitudes as the bond lengths change \cite{Oliva2013,OlivaLeyva2015,Naumis2017}.   
Such effects are easy to understand by thinking about the uniform expansion of a sheet of graphene. As all distances between Carbons atoms are uniformly lengthened, the strained reciprocal lattice basis vectors are  scaled by a constant factor with respect to the pristine graphene case. This is the geometrical effect. Simultaneously, the tight-binding hopping integral is also reduced by a constant factor leading to a change in the Fermi velocity; this is the energetic change. To further clarify such effects, imagine for a moment that we keep the distances between the atoms the same but change the tight-binding hopping integral parameter between neighbors. The Hamiltonian matrix will be modified while the reciprocal lattice remains the same. This results in a pure energetic effect as the eigenvalues of the Hamiltonian matrix change. A pure geometrical effect is achieved by keeping the hopping integral parameter the same but changing the distances between atoms. The Hamiltonian matrix remains the same while the reciprocal lattice is distorted. Thus, the eigenvalues of the Hamiltonian are the same as those in the unstrained system, but the energy dispersion gets distorted due to a different reciprocal lattice.    

The extra shift produced by the hoppings can be thought as a pseudomagnetic field in the low-energy Hamiltonian~\citep{GuineaStrain,Oliva2013,Nam2017}. In each layer, the pseudomagnetic vector potential, $\boldsymbol{A}^{(l)}=(A_{x}^{(l)},A_{y}^{(l)})$ is given by~\citep{GuineaStrain} 
\begin{align}
A_{x}^{(l)} & =\frac{\sqrt{3}}{2a}\beta_{G}[\varepsilon_{xx}^{(l)}-\varepsilon_{yy}^{(l)}],\nonumber \\
A_{y}^{(l)} & =\frac{\sqrt{3}}{2a}\beta_{G}[-2\varepsilon_{xy}^{(l)}],\label{eq:VectorPotGrunn}
\end{align}
where $\beta_{G}\approx3.14$ is the Grunesien dimensionless parameter. By taking into account the geometrical and energetic shifts, the final positions of the Dirac points are \cite{Oliva2013,OlivaLeyva2015,Naumis2017}
\begin{equation}
   \bm{D}_{\zeta}^{(l)}=\tilde{K}_{\bm \zeta}^{(l)}-\zeta \bm{A}^{(l)}=(\mathbb{I}-[\varepsilon^{(l)}]^{T})\boldsymbol{K}^{(l)}_{\zeta}-\zeta \bm{A}^{(l)} .
\end{equation}
In TBG with a small twist, the moir\'e superlattice constant is much larger than the atomic scale, and the low-energy physics is dominated by states near  $\boldsymbol{K}_{+}$ and $\boldsymbol{K}_{-}$; therefore, we can analyze
each valley separately. Thus, the low-energy Hamiltonian for TBG with uniaxial strain can be written as
\begin{equation}
H_\text{TBG}=\left(\begin{array}{cc}
H(\boldsymbol{q}_{\zeta}^{(1)}) & U^{\dagger}\\
U & H(\boldsymbol{q}_{\zeta}^{(2)})
\end{array}\right),\label{eq: HamiltonianStrained}
\end{equation}
where $\boldsymbol{q}_{\zeta}^{(l)}=R(\pm\theta/2)(\mathbb{I}+[\varepsilon^{(l)}]^{T})(\boldsymbol{q}-\bm{D}_{\zeta}^{(l)})$
with $\pm$ for $l=1$ and $l=2$, respectively.  $H(\boldsymbol{q})=-\hbar v_{F}\boldsymbol{q}\cdot(\zeta\sigma_{x},\sigma_{y})$ is the Hamiltonian for a monolayer. In the above equation, $U$ is the interlayer coupling between twisted graphene layers given by the Fourier expansion, 
\begin{align}
U & =\left(\begin{array}{cc}
u & u^{\prime}\\
u^{\prime} & u
\end{array}\right)+\left(\begin{array}{cc}
u & u^{\prime}\omega^{-\zeta}\\
u^{\prime}\omega^{\zeta} & u
\end{array}\right)e^{i\zeta\boldsymbol{g}_{1}\cdot\boldsymbol{r}}\nonumber 
  +\left(\begin{array}{cc}
u & u^{\prime}\omega^{\zeta}\\
u^{\prime}\omega^{-\zeta} & u
\end{array}\right)e^{i\zeta(\boldsymbol{g}_{1}+\boldsymbol{g}_{2})\cdot\boldsymbol{r}}
\label{eq: InterlayerU}
\end{align}
where $\omega=e^{2\pi i/3}$, with $u=0.0797$ eV and $u^{\prime}=0.0975$~eV~\citep{Koshino2018a} being the amplitudes which take into account out-of-plane corrugation effects~\citep{Koshino2018a,TKV19,Nam2017}.  As $U$ is affected by strain, it is expanded using the {\it moir\'e strained vectors} $\boldsymbol{g}_{i}$.

\section{Self-consistent Hartree Interaction for strained TBG/hBN}

The electron-electron interaction is introduced with a self-consistent Hartree potential in the TBG/hBN Hamiltonian 
\begin{equation}
 H=H_\text{TBG}+V_\text{hBN}+V_\text{H}
 \label{eq: MainHamil},
\end{equation}
where the first term is the usual strained TBG Hamiltonian, $V_\text{hBN}$ captures the effect of hBN on the adjacent graphene layer, and $V_\text{H}$ is the Hartree potential. Notice that we are considering a nearly aligned situation where the moire vectors of TBG are the same as those of hBN and its adjacent graphene layer. This condition is satisfied when $|\theta_\text{hBN}|\approx|\theta/2|$ with $\theta_\text{hBN}$ the twist angle between hBN and graphene and $\theta$ the TBG twist angle (see Ref.~\citep{CeaPantaGuinea2020} and Ref.~\citep{LongJose2021} for further details). The potential induced by the hBN layer on TBG is then given by~\cite{Wallbank2013,San-Jose2014}
\begin{eqnarray}
V_\text{hBN}\left(\boldsymbol{r}\right)= \omega_0 \sigma_{0}+
\Delta\sigma_{3}+
\sum_{j}
V_\text{SL}(\boldsymbol{g_j})e^{i\boldsymbol{g_j}\cdot\boldsymbol{r}},
 \label{eq: VSlhbN}
\end{eqnarray}
where $w_{0}$ and $\Delta$ represent a spatially-uniform scalar and mass term respectively. 
$\boldsymbol{g}_j$ runs over the six first vectors or first star of reciprocal lattice vectors. In the unstrained lattice, these six reciprocal lattice vectors satisfy $|\boldsymbol{b}|=4\pi/(\sqrt{3}L_m)$, where $L_{m}$ is the  moir\'e lattice constant. In the strained lattice, they are transformed according to Eq.~\eqref{eq: StrainRealPrimitive} and Eq.~\eqref{eq: Straingvectors}. We ignore contributions of smaller wavelengths in $V_\text{SL}$. The amplitudes $V_\text{SL}(\boldsymbol{g_j})$ are given by
\begin{equation}
V_\text{SL}(\boldsymbol{g_j})= \left[V_{s}^{e}+i(-1)^{j}V_{s}^{o}\right]\sigma_{0}+\left[V_{\Delta}^{o}+i(-1)^{j}V_{\Delta}^{e}\right]\sigma_{3}+\left[V_{g}^{e}+i(-1)^{j}V_{g}^{o}\right]M_j,
\label{eq: hbN Perturbation}  
\end{equation}
with $M_j=(-i\sigma_{2}g_{j}^{x}+i\sigma_{1}g_{j}^{y})/|\boldsymbol g_{j}|$. The parameters $V_{s}^{e}$ and $V_{s}^{o}$ are position-dependent scalar terms and are even and odd under spatial inversion, respectively. Similarly, $V_{\Delta}^{o(e)}$ and $V_{g}^{o(e)}$ are position-dependent mass and gauge terms, respectively. We use the parametrization of $V_\text{hBN}$ given in Ref.~\cite{LongJose2021},  
\begin{align}
(w_{0},\Delta,V_{s}^{e},V_{s}^{o},V_{\Delta}^{e},V_{\Delta}^{o},V_{g}^{e},V_{g}^{o})= \alpha
(0, 31.62, -0.75, 0.68,-0.02, 3.4, -5.14, 18.6) \text{ meV}
\label{eq: hbN parameters}  
\end{align}
The parameter $\alpha$ is introduced to take into account large twist angles, where the strain fields and the uniform mass gap dominate. In addition, it is important to mention that the first-star approximation for the effect of hBN  with the above parameterization can be used only for small strain values $\sim 1-5\%$. If the strain magnitude is larger, additional strain fields and even additional harmonics should be taken into account. To date, only Ref.~\citep{LongJose2021} has reported the full set of parameters for unstrained TBG on hBN. 

The Hartree interaction is introduced in the TBG/hBN Hamiltonian as a filling-dependent scalar potential parameterized by  $\rho_\text{H}$, a complex quantity encoding the charge density that is calculated self-consistently~\citep{Cea2019}. We approximate the Hartree potential by summing the first star of Fourier harmonics. In the unstrained lattice,  these Fourier harmonics are the six reciprocal lattice vectors satisfying $|\boldsymbol{b}|=4\pi/(\sqrt{3}L_m)$, where $L_{m}$ is the  moir\'e lattice constant. In the presence of uniaxial heterostrain, $\mathcal{C}_{3z}$ symmetry is broken, and the Fourier expansion of the Hartree potential is now given by three numbers $\rho_\text{H}(\boldsymbol{g}_1),$ $\rho_\text{H}(\boldsymbol{g}_2),$  and $\rho_\text{H}( -\boldsymbol{g}_1-\boldsymbol{g}_2)$, where $\boldsymbol{g}_1 \text{ and } \boldsymbol{g}_2$  are the reciprocal lattice vectors of the distorted triangular superlattice. Following Ref.~\citep{CeaPantaGuinea2020}, the matrix elements of the self-consistent Hartree potential are given by~\citep{Guinea2018a,Cea2019}
\begin{equation}
\rho_\text{H}(\boldsymbol g)=4V_{0}(\boldsymbol g) \int \frac{d^{2}\boldsymbol{k}}{V_\text{mBZ}}\sum_{\boldsymbol g^{\prime},l}\phi_{k,l}^{\dag}(\boldsymbol g^{\prime})\phi_{k,l}(\boldsymbol g+\boldsymbol g^{\prime}),\label{eq: rho components} 
\end{equation}
where $V_{0}(\boldsymbol g)=v_C(\boldsymbol g)/A_{c}$, with $v_C(\boldsymbol g)=2 \pi e^2/(\epsilon |\boldsymbol g|)$ is the Fourier transform of the Coulomb potential evaluated at $\boldsymbol g,$  $A_{c}$ is the area of the strained unit cell, and the factor 4 takes into account spin/valley degeneracy. Notice that the reciprocal lattice vectors $\boldsymbol{g}_{i}$ are those of the strained lattice. Therefore, the Fourier expansion of the Hartree potential in real space can be written as 
\begin{equation}
V_\text{H}(\boldsymbol{r})=2\sum_{i=1}^3\rho_\text{H}(\boldsymbol g_{i})\cos(\boldsymbol g_{i}\cdot\boldsymbol{r}). 
\label{eq: VhRealSpace}
\end{equation}
In pristine TBG, the matrix elements of the Hartree potential satisfy $\rho_\text{H}(\boldsymbol g_{1})=\rho_\text{H}(\boldsymbol g_{2})=\rho_\text{H}(\boldsymbol g_{3})$ due to $C_{3z}$ symmetry. In the presence of strain, the inequivalence of the reciprocal vectors in the triangular superlattice implies that $\rho_\text{H}(\boldsymbol g_{1})\neq\rho_\text{H}(\boldsymbol g_{2})\neq\rho_\text{H}(\boldsymbol g_{3}),$ which is consistent with the broken $C_{3z}$ symmetry in the strained system. In the presence of an hBN substrate, the mass terms break the remaining $C_{2z}$ symmetry. Notice that Eq.~\eqref{eq: VhRealSpace} has the form of a periodic scalar potential. The Hartree potential is obtained by self-consistently solving for the three parameters $\boldsymbol \rho_\text{H}(\boldsymbol g_i).$ To solve the self-consistent Hartree Hamiltonian, the charge distribution is approximated as $\rho_\text{H}=\rho_{0}+\delta\rho_\text{H},$ where $\rho_{0}$ is a constant which takes into account the total density from all bands not included in the calculations~\citep{Guinea2018a}. The charge distribution is fixed by assuming a homogeneous state at charge neutrality. Therefore, the integral in Eq.~\eqref{eq: rho components} is evaluated only over energy levels with $\epsilon_n(\boldsymbol k)$ between half filling and Fermi level, $\mu$.  For a given value of the Fermi level, we calculate the miniband spectrum and wavefunctions with self-consistent diagonalization of the Hamiltonian by considering up to $N=91$ vectors in  reciprocal space.

\section{Nonlinear Hall Effect}

 Following Refs.~\citep{Du2019,du2020quantum,TonyLowPaco2015,SodemannLiang2015}, the transverse currents up to second order in the electric field are $\mathcal{J}_{a,\mu}=\text{Re}\{\mathcal{J}_{a,\mu}^0+\mathcal{J}_{a,\mu}^\omega e^{i\omega t}+\mathcal{J}_{a,\mu}^{2\omega} e^{i2\omega t}\}$ with a rectified component $\mathcal{J}_{a,\mu}^0=\chi_{abc}^{\mu}\xi_{b}\xi_{c}^*$, a linear term $\mathcal{J}_{a,\mu}^\omega=\sigma_{ab}^{\mu}\xi_{b}$, and a second-harmonic component $\mathcal{J}_{a,\mu}^{2\omega}=\chi_{abc}^{\mu}\xi_{b} \xi_{c}$, with susceptibilities given by
\begin{eqnarray}
\sigma_{ab}^{\mu} & = & -\frac{e^{2}}{\hbar}\sum_{n}\int\frac{d\boldsymbol{k}}{(2\pi)^{2}}\varepsilon^{abc}f_{0,\mu}(\boldsymbol{k})\Omega_{n,\mu}^{c}, \label{eq: Conductbulk} \\
\mathcal{\chi}_{abc}^\mu & = & -\frac{e^{3}\tau}{2\hbar^{2}(1+i\omega\tau)}\sum_{n}\int\frac{d\boldsymbol{k}}{(2\pi)^{2}}\varepsilon^{acd}\Omega_{n,\mu}^{d}\partial_{b}f_{0}(\boldsymbol{k}),
\label{eq: Conductsurf}
\end{eqnarray}
where $f_{0,\mu}=f_{0}[\epsilon_{n,\mu}(\boldsymbol{k})]$ is the equilibrium Fermi-Dirac distribution, $\epsilon_{n,\mu}(\boldsymbol{k})=\epsilon_{n}(\boldsymbol{k})-\mu$ with $\epsilon_{n}(\boldsymbol{k})$ the energy of the band and $\mu$  the Fermi energy, and $\tau$ is the scattering time. Equation~\eqref{eq: Conductbulk} is the linear contribution to the Hall current; all states below chemical potential contribute to this term. The second term, Eq.~\eqref{eq: Conductsurf}, is responsible for the nonlinear Hall effect, wherein only states close to the Fermi surface contribute. From Eq.~\eqref{eq: Conductsurf}, we can isolate the Berry dipole as
\begin{equation}
D_{bd,\mu}=\sum_{n}\int \frac{d\boldsymbol{k}}{(2\pi)^{2}} \Omega_{n,\mu}^{d}(\boldsymbol{k})\partial_{b}f_{0}[\epsilon_{n,\mu}(\boldsymbol{k})]
\label{eq: Berry Dipole}.
\end{equation}
Therefore, the susceptibility in Eq.~\eqref{eq:  Conductsurf}  can be rewritten more compactly as 
\begin{equation}
\mathcal{\chi}_{abc}^\mu  =  -\frac{e^{3}\tau}{2\hbar^{2}}\varepsilon^{acd}D_{bd,\mu},
\label{eq: SurfDipoleTotal}
\end{equation}
where we are assuming that $\omega \tau \ll1 $, because $\omega$ is about tens of Hertz and the relaxation time $\tau$ is about picoseconds in experiments~\citep{Brida2013Time}. For an electric field that makes an angle $\gamma$ with the $x$-axis (which is defined by the lattice), we can write $\boldsymbol{\xi} = \xi_0\left(\cos \gamma,\sin \gamma \right),$ where $\xi_0$ is a real number. $\boldsymbol{\xi} = \left(\xi_x,\xi_y \right)$ is in general a two-dimensional vector of complex numbers, which describes an electric field of arbitrary polarization. However, we only consider linear polarization where $\boldsymbol{\xi}$ is real and defines a definite direction. In this case, $\mathcal{J}_{a,\mu}^0 = \mathcal{J}_{a,\mu}^{2\omega};$ so we only focus on $\mathcal{J}_{a,\mu}^0.$ Its components are given by 
\begin{equation}
    \begin{split}
        \mathcal{J}_{x,\mu}^0 &= \chi^\mu_{xxy} \xi_x\xi_y +\chi^\mu_{xyy} \xi_y \xi_y = \xi_0^2 \sin \gamma \left[ \chi^\mu_{xxy} \cos \gamma +\chi^\mu_{xyy} \sin \gamma  \right],\\
        \mathcal{J}_{y,\mu}^0 &= \chi^\mu_{yxx} \xi_x\xi_x +\chi^\mu_{yyx} \xi_y \xi_x = - \xi_0^2 \cos \gamma  \left[ \chi^\mu_{xxy} \cos \gamma +\chi^\mu_{xyy} \sin \gamma  \right],
    \end{split}
\end{equation}
where we have used $\chi^\mu_{xxy} = -\chi^\mu_{yxx}$ and $\chi^\mu_{yyx} = -\chi^\mu_{xyy}.$ This motivates us to define an effective angular susceptibility 
\begin{equation}
    \chi_\text{eff}^\mu \left( \gamma \right) = \chi^\mu_{xxy} \cos \gamma +\chi^\mu_{xyy} \sin \gamma 
\end{equation}
using which the current can be written simply as $\boldsymbol{\mathcal{J}}^0 = \xi_0^2 \chi_\text{eff}^\mu (\gamma) \left( \sin\gamma, - \cos \gamma \right).$ We note that $\boldsymbol{\mathcal{J}}^0 \cdot \boldsymbol{\xi} = 0$ as required for transverse currents. This form of the susceptibility is especially useful for comparison with experimental data. If an electric field of magnitude $\xi_0$ is applied at an angle $\gamma$ with respect to the crystallographic axes, then a current of magnitude $\xi_0^2 \chi_\text{eff}^\mu$ develops in the direction that makes an angle $\gamma - 90^\circ$ with the same axes. Finally, we multiply the effective susceptibility by $4$ to account for spin and valley degeneracies
\begin{equation}
\begin{split}
     \chi_\text{eff}^\mu \left( \gamma \right) &= 4\chi^\mu_{xxy} \cos \gamma +4\chi^\mu_{xyy} \sin \gamma   \\
     &= -\frac{4e^3 \tau}{2\hbar^2} \left[ D_{xz,\mu} \cos \gamma + D_{yz,\mu} \sin \gamma\right] = -\frac{e^3 \tau}{2\hbar^2} D_{\text{eff},\mu}(\gamma),
\end{split}
\end{equation}
where we have defined an effective angular Berry dipole
\begin{equation}
    D_{\text{eff},\mu}(\gamma) = 4 \left[ D_{xz,\mu} \cos \gamma + D_{yz,\mu} \sin \gamma\right] = 4 \boldsymbol{D}_\mu \cdot \hat{\boldsymbol{\xi}}.
\end{equation}
Apart from the factor of $4,$ this is nothing more than the Berry dipole along the direction of the electric field.

\section{Estimation of the Hall Voltage using the Berry dipole}

The Berry dipole in TBG/hBN has two components $D_{xz}$ and $D_{yz}$. In the following, to numerically compare our results with the experiment in Ref. \cite{Duan2022Nonlin}, we calculate the transverse Hall voltage $V_\text{H}^{2\omega}$ as a function of an external electric current $I^\omega$. As described in the previous section, the current density is given by
\begin{equation}
\mathcal{J}^{2\omega} = \xi_0^2 \chi_\text{eff}^\mu (\gamma) \left( \sin\gamma, - \cos \gamma \right).
\end{equation}
The electric field and the current density can be rewritten in terms of the electric current and the Hall voltage by using $\xi_0=I_0/(\sigma L)$ and $\mathcal{J}^{2\omega}=\sigma V_\text{H}^{2\omega}/W$ respectively. Here, $\sigma_d$ is the conductivity which is assumed to be isotropic, and $L$ and $W$ are the length and width of the sample, respectively. Assuming $W \approx L$ for simplicity, we can write the Hall voltage as        
\begin{equation}
V_\text{H}^{2\omega} =-\frac{e^{3}\tau}{2\hbar \sigma_{d}^{3} W} D_{\text{eff},\mu}(\gamma) I_{0}^{2}, 
\end{equation}
and from the Drude formula $\sigma_d = ne^{2}\tau/m^{*}$, the above equation takes the form
\begin{equation}
V_\text{H}^{2\omega}=-\frac{e m^{*}}{2\hbar^{2}n \sigma_{d}^2 W}D_{\text{eff},\mu}(\gamma) I_{0}^2. 
\label{eq: Vhall}
\end{equation}
Using the parameters: $m^* = 3 m_e$ with $m_e$ the mass of electron~\cite{Cao2018_bis}, electronic density as a function of filling $\nu = 4 n/n_s$, with $n_s= \num{2.5e12} \text{ cm}^{-2}$, sample width $W\sim 1.5$ $\mu m$~\cite{Sinha2022diposenses}, conductivity $\sigma_d \sim \num{1e-4} \text{ S},$ and electric current as in Ref.~\cite{Duan2022Nonlin} of $I_0\sim 1$ $\mu A$, a Hall voltage at $\nu=-4$ with magnitude $|V_\text{H}^{2\omega}|\sim 1000$ $\mu V$ is obtained for $\gamma=45^{\circ}$ and $D_{xz}=D_{yz}=3.5$ nm. These estimates provide reasonable order-of-magnitude agreement with the experiment. However, one should approach these estimates with skepticism since they do not capture many complications inevitably present in an actual experiment. For instance, the conductivity, the effective mass, and the charge density may vary as a function of the chemical potential and  strongly depend on experimental conditions. These complexities notwithstanding, our results suggest that the Berry dipole required to obtain the Hall voltages reported in Ref.~\cite{Duan2022Nonlin} is similar in order of magnitude to our calculations. 

\section{Additional results: TBG with hBN}
\begin{figure*}
\begin{centering}
\includegraphics[scale=0.53]{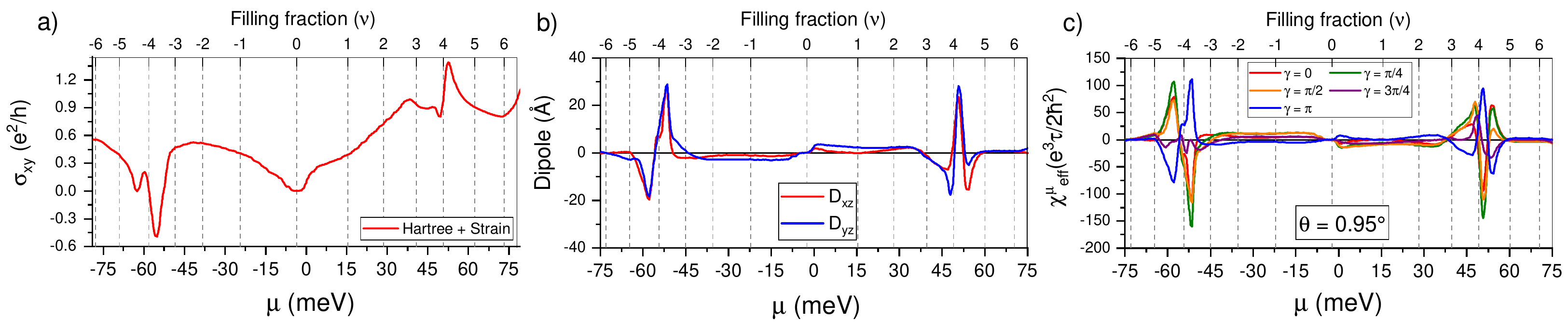}
\par\end{centering}
\caption{Interaction-enhanced topological Hall effects in strained ($\epsilon = 0.2\%$ and $\phi = 0$) TBG/hBN for $\theta = 0.95^\circ$ and $\alpha = 0.50$ in Eq.~\eqref{eq: hbN parameters}. In (a), we show the linear Hall conductivity. In (b), the large Berry dipole is enhanced near $\pm 4$. In (c), the effective Hall susceptibility is strongly enhanced depending on the current direction. 
\label{fig: FigureShBN}}
\end{figure*}

In this section, we consider the case when the potentials induced by the hBN in TBG are smaller. This occurs when hBN and graphene are misaligned by a larger relative twist. Recently~\cite{LongJose2021}, it has been shown that the effects of the periodic potentials are relevant if the angle between hBN and the lower graphene layer, $|\theta_\text{hBN}|$, is between $0$ and $ \sim 3^\circ$; on the other hand,  main contribution at larger angles comes from the strain fields and the uniform mass gaps. To take into account samples with a different degree of alignment with hBN,  we display in Fig.~\ref{fig: FigureShBN} the topological Hall effects in strained TBG/ hBN for $\theta = 0.95^\circ$ and $\alpha = 0.50$ in Eq.~\eqref{eq: hbN parameters}. This is a reasonable approximation for a twist angle  $|\theta_\text{hBN}|$ between $1.5^\circ$ and $2.0^\circ$ where the mass term is reduced by a half of its magnitude. 

\section{Band structures: strained and unstrained systems}
\begin{figure*}
\begin{centering}
\includegraphics[scale=0.35]{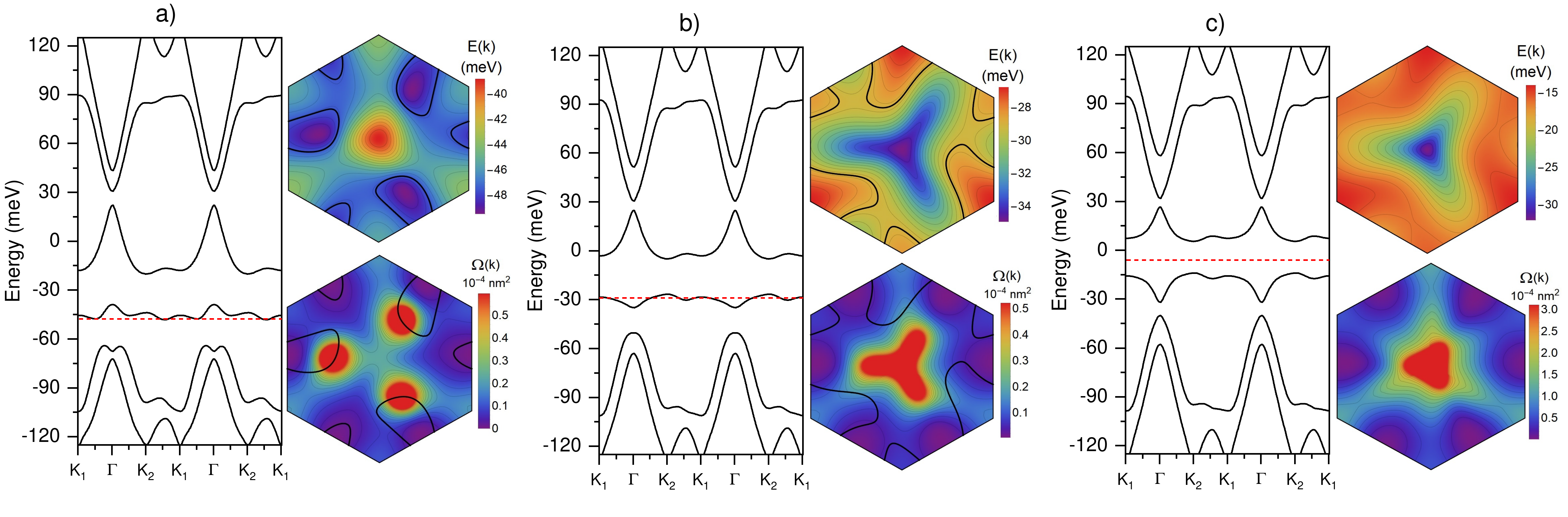}
\par\end{centering}
\caption{Band structure of unstrained TBG nearly aligned with hBN as a function of the chemical potential for a filling factor: (a) $\nu=-3$, (b) $\nu=-1,$ and (c) $\nu=0$. The red dashed line in each figure is the Fermi level. We also show density plots for the active bands. The top panels are the energy bands, and bottom panels are the Berry curvatures. The black lines in each density plot trace out the corresponding Fermi surfaces.   
\label{fig: FigureSa}}
\end{figure*}

\begin{figure*}
\begin{centering}
\includegraphics[scale=0.35]{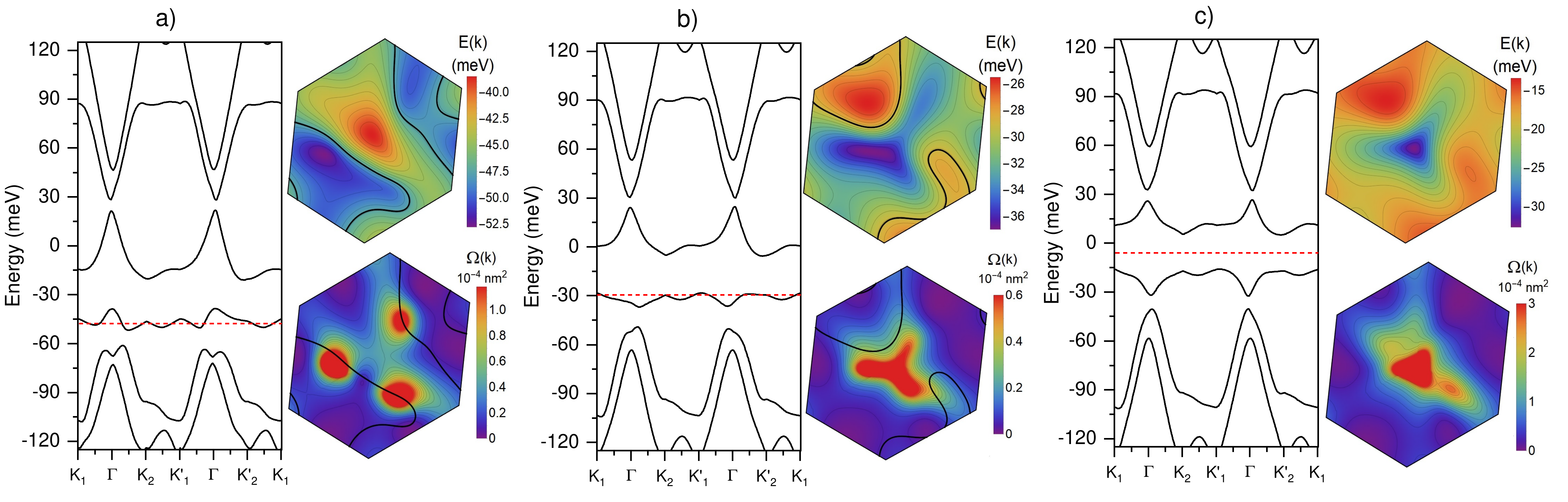}
\par\end{centering}
\caption{Band structure of strained ($\epsilon=0.2 \%$ and $ \phi=0$)  TBG nearly aligned with hBN as a function of the chemical potential for a filling factor: (a) $\nu=-3$, (b) $\nu=-1,$ and (c) $\nu=0$. Red dashed line in each figure is the Fermi level. We also show density plots for the active bands. The top panels are the energy bands, and bottom panels are the Berry curvatures. The black lines in each density plot trace out the corresponding Fermi surfaces.   
\label{fig: FigureSb}}
\end{figure*}

In Fig.~\ref{fig: FigureSa}, we display the band structure of unstrained TBG nearly aligned with hBN as a function of the chemical potential for different filling factors. It is clear that both energy spectrum and topology are  modified as a function of filling. Band structures and distribution of the Berry curvature within the mBZ are mapped out in the hexagonal panels. In Fig.~\ref{fig: FigureSb}, we introduce a uniaxial strain $\epsilon = 2\%$ and $\phi = 0$. As in the previous case, the band structure and topology are also modified. However, the presence of uniaxial heterostrain and hBN allow for a non-zero Berry dipole. Notice that the states close to the Fermi boundary (black lines) are those contributing most to the Berry dipole. 

\section{Valley Hall conductivity for a non interacting TBG/hBN}

\begin{figure*}
\begin{centering}
\includegraphics[scale=0.44]{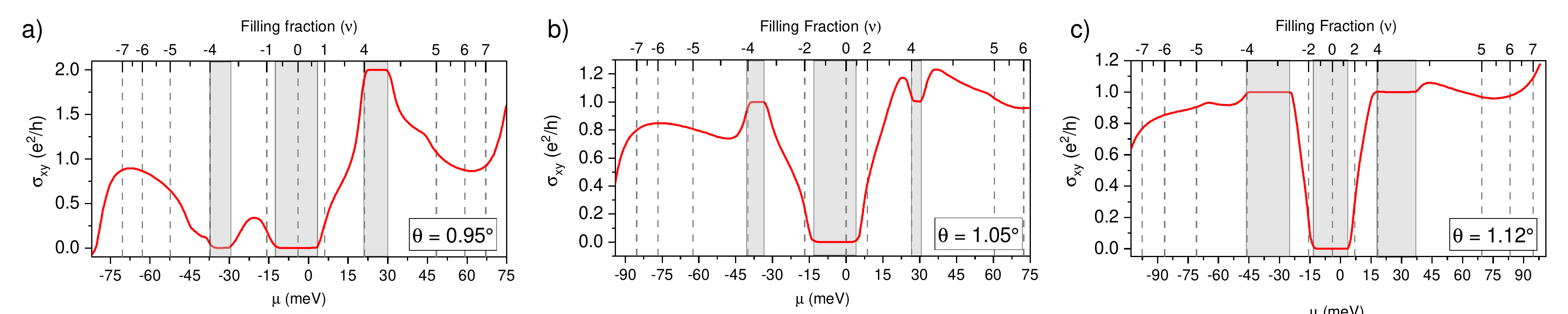}
\par\end{centering}
\caption{Valley Hall conductivity for the unstrained and non-interacting TBG/hBN system. Different  twist angles are indicated. Shaded gray regions are the gaps where the conductivity is quantized. 
\label{fig: FigureSc}}
\end{figure*}

In Fig.~\ref{fig: FigureSc}, we show the valley Hall conductivity of  TBG/hBN in absence of both strain and Hartree potential. 

\section{Numerical evaluation of the Hall conductivities}

In this section, we address the numerical evaluation of the linear and nonlinear Hall terms in Eq.~\eqref{eq: Conductbulk} and Eq.~\eqref{eq: Conductsurf}. In numerical evaluation, we first define an hexagonal grid centered at $\Gamma$ within the mBZ of Fig. 1(c). In our calculations, we use a grid size of $3N^{2}$ $\boldsymbol{k}$-points, with $N\sim20-30$ for the self-consistent Hartree calculations (see Ref.~\citep{Cea2019} for further details) and $N\sim300-400$ for the Berry curvature. We diagonalize Eq.~\eqref{eq: MainHamil} and from its eigenvectors, we obtain $\Omega_{n,\mu}(\boldsymbol{k})$ following the procedure described in Ref.~\citep{Fukui2005}. By writting
\begin{equation}
\boldsymbol{\Omega}_{n,\mu}(\boldsymbol{k})=\nabla_{\boldsymbol{k}}\times\boldsymbol{A}_{n,\mu}(\boldsymbol{k}),\:\boldsymbol{A}_{n,\mu}(\boldsymbol{k})=-i\left\langle \Psi_{nk,\mu}\right|\nabla_{\boldsymbol{k}}|\left. \Psi_{nk,\mu}\right\rangle ,
\label{eq:Omega}
\end{equation}
we numerically integrate the Berry connection  $\boldsymbol{A}_{n,\mu}(\boldsymbol{k})$ in small loops around each momentum $\boldsymbol{k}$. For each loop, we choose a set of eigenvectors around the loop and then we calculate the Berry connection between points of the loop. The total contribution of each small loop is the local Berry curvature determined up to a prefactor. The sum of these loops over the Brillouin zone gives the Chern number of the corresponding band modulo $2\pi$. The integrals in Eq.~\eqref{eq: Conductbulk} and Eq.~\eqref{eq: Conductsurf} are obtained by summing the Berry curvature at each point weighted by the corresponding scalar function at  Fermi level $\mu$ and a given temperature. This numerical procedure allows us to eliminate  problems with gauge ambiguity because the arbitrary phases appear twice with opposite sign. However, depending on the distribution of the Berry curvature within the mBZ and the temperature, the size of the loop must be small enough to achieve convergent numerical results. 
\end{document}